# Dynamical screening effects on formation of swift heavy ions damage in GaN

S.V. Moskalets[1], R.A. Rymzhanov[2,3], A.E. Volkov[1]

[1]*P.N. Lebedev Institute of the Russian Academy of Sciences, Leninskij pr., 53,119991 Moscow, Russia*

[2]*Flerov Laboratory of Nuclear Research, Joint Institute for Nuclear Research, Joliot-Curie, 6, 141980 Dubna, Russia*

[3]*The Institute of Nuclear Physics, Ibragimov St. 1, 050032, Almaty, Kazakhstan*

## Abstract

We used a multiscale model to study the damage caused by swift heavy ions in GaN. The model combines Monte Carlo and molecular dynamics simulations to analyze the material response to the excitation initiated by a projectile. We found that the most appropriate simulation approach couples the dynamical screening of charges of target atoms during the scattering of fast electrons with the Tersoff-Brenner interatomic potential describing atomic dynamics in the excited target. The simulations demonstrate the formation of damaged ion tracks with an amorphous core, containing voids filled with nitrogen. The core is surrounded by a damaged crystalline region containing edge dislocations, consistent with experimental observations.

## 1. Introduction

Gallium nitride (GaN) is a wide-bandgap (3.4 eV) semiconductor widely used in modern electronic devices [1]. GaN-based field-effect transistors are highly resistant to radiation making GaN a promising material for space technologies [2]. In deep space missions, galactic cosmic rays may cause single event effects or long-term material degradation of electronic components. Currently, the effect of the heavy component of galactic cosmic rays on GaN has not been well studied, which can be effectively addressed by investigating the results of irradiation with swift heavy ions (SHI) with masses $> 10\ m_p$ and energies ~ 1-10 MeV/nucleon.

Refs. [3-6] demonstrated that irradiation with SHIs leads to the formation of damaged ion tracks in crystalline GaN (α-phase). These tracks appear in gallium nitride when the ion electronic energy losses exceed the threshold of 15 keV/nm. Ref. [4] discovered that irradiation of GaN with ions just above the threshold of the electronic stopping results in the creation of discontinuous tracks. Ions with higher energies produce cylindrical tracks with continuous amorphous cores. These results were confirmed in [5, 6], where additionally cavity formation in subsurface regions along the ion trajectory was detected. However, these experiments did not provide detailed information about the structure of the defective region surrounding the track core. The microscopic kinetics and mechanisms of track formation also remain elusive in current experiments.

Numerical modeling can improve knowledge about SHI track formation in GaN. Advanced simulations in [5, 6] have demonstrated the generation of nitrogen voids and dislocations along the SHI track in GaN. Regrettably, the authors relied on the two-temperature thermal spike model (TTM) combined with molecular dynamics to describe the excitation of the electronic and atomic systems of the material [7]. Despite its popularity, TTM has limited predictive power [8].

We apply the multiscale TREKIS+MD [8] model to describe SHI track formation in GaN. Taking into account the significantly different response times of electronic and atomic systems, the approach divides the track kinetics into two stages. The Monte Carlo (MC) code TREKIS-3 (Time Resolved Electron Kinetics in SHI-Irradiated Solids) describes the excitation and relaxation of the target electronic system, as well as energy transfer to the atomic system [9, 10]. The spatial distribution of the transferred energy into the lattice at ~100 fs after the projectile passage is then used as the initial condition in the classical molecular dynamics (MD) code LAMMPS (Large-scale Atomic/Molecular Massively Parallel Simulator) [11] to describe the further relaxation of the atomic system. This coupling of the approaches gives us the ability to monitor the evolution of the electronic and atomic systems in GaN at a microscopic level during the formation of the SHI track, from the ion passage to the cooling of the atomic system.

For the first time we analyze and demonstrate the importance of the dynamical screening of atomic charges on energy transfer from excited electrons to the atomic lattice, affecting final damage in SHI tracks in GaN. The simulation results represent the size and structure of the SHI track core as well as defects in the surrounding crystalline lattice.

## 2. Model

### *2.1 Monte-Carlo modeling*

The MC model TREKIS-3 [9] describes the electron kinetics considering ionization of atoms by an incident SHI and the generation of fast $\delta$-electrons, relaxation of deep shell holes, transport and scattering of secondary electrons, valence holes and photons, as well as energy transfer to the crystal lattice. The simulation spans 100 fs after the projectile passage, restricted by the cooling down of the electronic subsystem. The transport of SHI, electrons, and valence holes in the material is tracked as a sequence of scattering events defined by interaction cross section $\sigma$. It is assumed that the particles have free motion between scattering events and their trajectories can be considered linear. The model assumes the homogeneous atomic structure

In the first order Born approximation, the differential scattering cross section of fast projectiles on a coupled ensemble can be expressed via the dynamic structure factor (DSF, $S(q,\omega)$) [12], taking into account the collective response of the target to the excitation by the incident particle. Due to the fluctuation-dissipation theorem [13], the DSF of the system can be expressed through the energy loss function (ELF) $Im\left(-\frac{1}{\varepsilon(q,\omega)}\right)$, where $\varepsilon(q,\omega)$ is the complex dielectric function (CDF) of the material. This gives the following form of the cross-section for scattering of charged projectiles [8, 14]:

$$\frac{d^2\sigma}{d(\hbar q)d(\hbar\omega)} = \frac{2Z_{in}^2(v,q)Z_t^2(v,q)}{n_{sc}\pi\hbar^2 v^2}\frac{e^2}{\hbar q}\left(1-e^{-\frac{\hbar\omega}{k_b T}}\right)^{-1} Im\left(-\frac{1}{\varepsilon(q,\omega)}\right). \quad (1)$$

Here $\hbar\omega$ and $\hbar q$ are the energy and momentum, transferred from an incoming particle, $n_{sc}$ is the scattering centers concentration, $v$ is the incoming particle velocity, $Z_{in}$ and $Z_t$ are the effective charges of incoming and target particles respectively, $e$ is the electron charge, $\hbar$ is Planck's constant, $k_b$ is Boltzmann's constant, and $T$ is the target temperature. The loss function $Im\left(-\frac{1}{\varepsilon(q,\omega)}\right)$ can be reconstructed from the optical data (see below).

Interaction of an incoming ion with a target causes changes in the ion charge. After a short projectile run, the Barkas model [15] can be used to determine the effective ion charge, reproducing the energy losses of a scattered particle within CDF-DSF formalism:

$$Z_{in}(v) = Z_{ion}\left(1 - exp\left(-\frac{v}{v_0}Z_{ion}^{-\frac{2}{3}}\right)\right), \qquad (2)$$

where $Z_{ion}$ is the SHI atomic number, and $v_0 = \frac{c}{125}$ is Barkas's correction to the Bohr velocity. In the case of electrons and valence holes $Z_{in}^2 = 1$.

Target ionization induces the appearance of valence and deep shell holes. The core holes decay via two channels: Auger or radiative decays. This causes formation of holes on higher atomic shells or the valence band, as well as the generation of electrons or photons. The characteristic times and probabilities of these decays were obtained from the EPICS-2017 database [16]. The TREKIS-3 also accounts for the movement of valence holes and their elastic scattering, which induce heating of the target atoms.

Extreme electronic excitation initiated by SHI alters the interatomic potential of a material, leading to the acceleration of atoms that raises their kinetic energy ("nonthermal lattice heating") [17]. Based on the effects of non-thermal bandgap collapse [8], this increase in the energy of atoms can be accounted for by the transformation of the potential energy of valence holes to the kinetic energy of the atomic ensemble at the end of the MC calculation (~100 fs after the ion passage) [8, 18, 19].

Thus, TREKIS-3 implements two channels for transferring energy from the excited electronic system to the atomic lattice: a "thermal" one, through the scattering of electrons and valence holes on atoms, and a "non-thermal" channel, through the transfer of the potential energy of electron-hole pairs.

*2.2 Effective charge models of target atoms*

During ionization events, the charge of a scattering center $Z_t$ is equal to the charge of the electrons being ionized ($Z_t = 1$). The generated electrons and holes participate in elastic scattering on target atoms. In these non-ionizing events, the effective charge of target atoms $Z_t$ is essential: the reaction of valence and atomic electrons to a projectile impact depends on the velocity of the incoming particle, which may affect the ability of the target electrons to screen the atomic nucleus during the scattering event [20].

In this work we consider three models describing the dependence of $Z_t$ on the velocity $v$ of the incident particle. The simplest case suggested in Ref. [21] uses $Z_t = 1$, which may not be appropriate in some cases because it does not account for screening effects.

The ionic core screening by target electrons can be approximated by the Barkas-like formula (BL) [22]:

$$Z_t(v) = 1 + (Z_{at} - 1)\left(1 - exp\left(-\frac{v}{v_0}(Z_{at} - 1)^{-\frac{2}{3}}\right)\right). \qquad (3)$$

This expression reduces to $Z_t = 1$ at low velocities of an incoming particle but takes into account the reduction of the screening effect for high velocities increasing the effective charge of a target atom $Z_t$ and resulting, finally, in scattering on the stripped atomic core.

Another approach accounting for dynamical screening (DS) is outlined in Ref. [20]. The authors break down the electronic part of the DSF into components that quantify the impact of valence (free) and bound electrons on the dynamic response of the electronic system of a target [23]. This leads to the following effective charge of a target atom:

$$Z_t(v, q) = Z_{at} - Z_I f_I(\tilde{q}) - \rho(\omega, \tilde{q}), \tag{4}$$

where $f_I$ is the atomic form-factor, taking into account only deep-shell electrons, $Z_I = Z_{at} - N_{VB}$, $N_{VB}$ is the number of electrons in the valence band of the material per atom, $\rho(\omega, \tilde{q})$ is the charge density of the free electron cloud around the atom, $\tilde{q} = \frac{q+q_{in}}{2}$ is an *ad hoc* parameter [20], accounting for the momentum of the incoming particle $q_{in}$ and the momentum, transferred into the atomic system, $q$. The larger $q_{in}$, the less the atomic screening effect is. Below we compare these three models considering the energy transferred to the atomic ensemble.

## *2.3 Reconstruction of the energy loss function*

Available optical refractive and absorption data can be used to reconstruct the dependence of the energy loss function (ELF) on the energy of a particle moving through a medium [24, 25]. The application of the Ritchie and Howie algorithm [26] allows fitting this experimental ELF as a sum of Lorentz-Drude oscillator functions:

$$Im\left(-\frac{1}{\varepsilon(q,\omega)}\right) = \sum_i \frac{A_i \gamma_i \hbar \omega}{\left(\hbar^2 \omega^2 - \left(E_{0i} + \frac{\hbar^2 q^2}{2m_e}\right)^2\right)^2 + (\gamma_i \hbar \omega)^2}. \tag{5}$$

The parameters $A_i$, $E_{0i}$, and $\gamma_i$ can be interpreted as the intensity, energy, and inverse lifetime of $i$-th collective excitation in the system. These values are determined through the fitting procedure.

The correctness of the obtained expression was verified using the sum rules [8]. The *f*-sum rule states that:

$$\frac{2}{\pi \Omega_p^2} \int_{I_p}^{\infty} Im\left(-\frac{1}{\varepsilon(q=0,\omega)}\right) \hbar \omega \, d(\hbar \omega) = N_e, \tag{6}$$

where $N_e$ is the number of electrons on a shell with ionization potential $I_p$, $\Omega_p = \sqrt[2]{\frac{4\pi n_m e^2}{m_0}}$ is the plasma frequency ($n_m$ is the density of GaN molecules (or formula units), $m_0$ is the free electron mass). The *KK*-sum rule states that the Kramers-Kronig dispersion relation of total ELF tends to unity:

$$\frac{2}{\pi}\int_0^\infty Im\left(-\frac{1}{\varepsilon(q=0,\omega)}\right)\frac{d(\hbar\omega)}{\hbar\omega} = 1. \tag{7}$$

Table 1 presents the parameters of GaN ELF and the assessment of fitting quality with sum rules. The calculated sum rules values are in good agreement with the expected ones (shown in parentheses), indicating a high quality of the fitting. Fig. 1 demonstrates ELF values reconstructed from the optical data and fitted ELF for GaN. The peaks in the graph correspond to deep atomic shells, the valence band, and optical phonons.

*Table 1. ELF fitting parameters and sum rules values*

|  | $E_{0i}$ | $A_i$ | $\gamma_i$ | *KK*-sum | *f*-sum ($N_e$) |
|---|---|---|---|---|---|
| Valence band (N + $M_{IV\,V}$) | 35.3 | 977 | 101 | 0.78504 | 17.97781 (18) |
|  | 18.9 | 116 | 5 |  |  |
|  | 23.7 | 24 | 3.24 |  |  |
|  | 13.8 | -18.7 | 7.7 |  |  |
|  | 7 | -11.6 | 8 |  |  |
| $M_{I\,II\,III}$-Ga subshell | 164 | 55 | 180 | 0.01202 | 8.03395 (8) |
| K-N shell | 320 | 316 | 150 | 0.00032 | 1.93287 (2) |
| L-Ga shell | 1270 | 600 | 680 | 0.00018 | 7.95068 (8) |
| K-Ga shell | 8600 | 230 | 6300 | 0 | 2.01809 (2) |
| Phonon peak | 9.16E-2 | 5.17E-4 | 3.64E-3 | 0.061445 | -- |
| Result: |  |  |  | 0.859005 (1) | 37.9134 (38) |

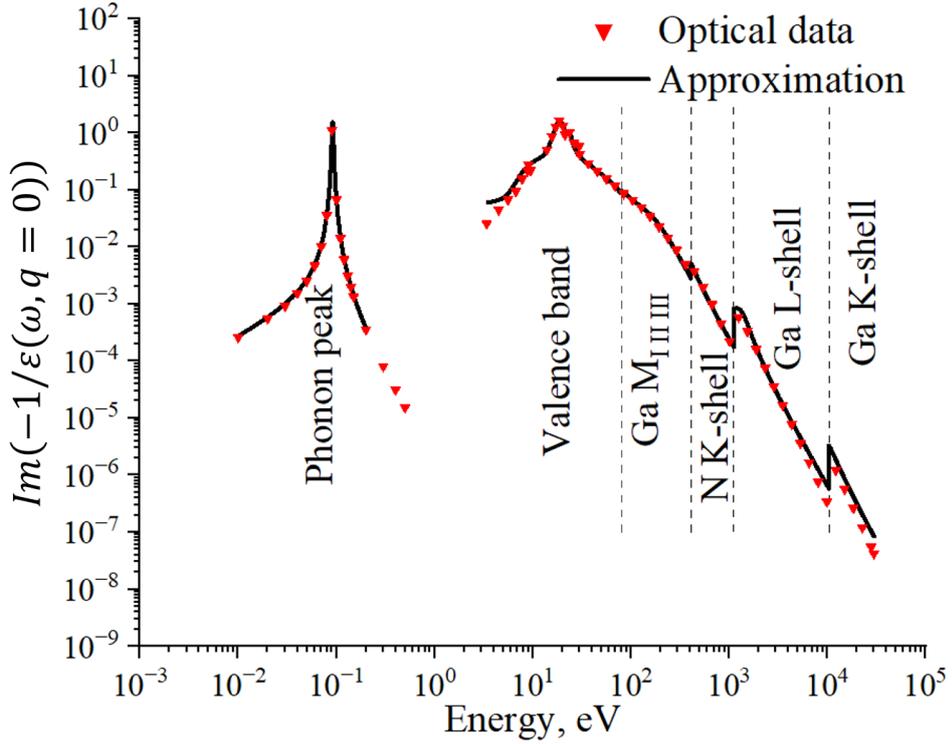

*Fig. 1 ELF of GaN reconstructed from optical data [24, 25] and its analytical approximation.*

## *2.4 Molecular dynamics modeling*

The initial positions of atoms match their arrangement in the unexcited crystal lattice of wurtzite GaN. The initial velocities of atoms are distributed throughout the material around the SHI trajectory according to the final spatial distribution of the kinetic energy of atoms calculated with TREKIS-3 [9, 10].

The relaxation of the atomic system was simulated in the LAMMPS program [11] and visualized using OVITO software [27]. The modeling of atomic dynamics in the SHI track spans approximately 100 picoseconds, corresponding to the full relaxation of the material lattice.

The structure of wurtzite GaN consists of primitive cells containing two gallium atoms and two nitrogen atoms. The cell parameters are $a = b = 3.19$ Å, $c = 5.19$ Å, $\alpha = \beta = 90°, \gamma = 120°$ [28]. Fig. 2 illustrates the unit cell of GaN and the hexagonal structure formed by these cells.

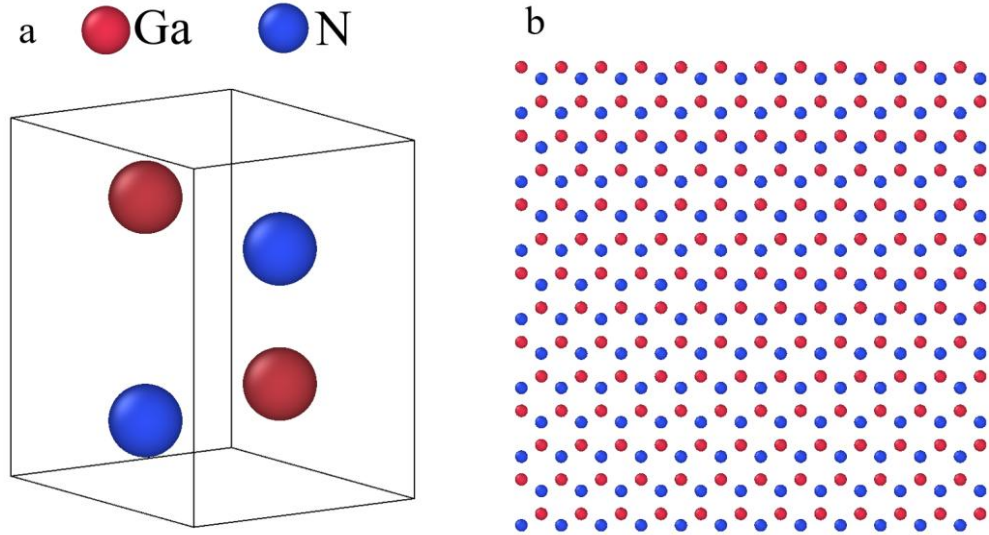

*Fig. 2 (a) Primitive cell of GaN and (b) projection of the structure along c-direction of the material.*

The orthogonal supercell used for SHI impact simulations was 25×25×25 nm³ in size. Periodic boundary conditions are applied, and the side walls parallel to the SHI trajectory are cooled to room temperature using the Berendsen thermostat [29] with damping time of 1000 fs, considering heat dissipation into the surrounding material. The energy distribution density obtained from MC modeling for various effective charges ($Z_t$ = 1, Barkas, DS) is utilized to assign initial velocities to the atoms.

We compared three interatomic potentials for the formation of damaged SHI tracks in gallium nitride: the Stillinger-Weber (SW) potential [30], which accounts for two-body and three-body interactions, the Tersoff-Brenner (TB) potential [31], which considers the bond strength dependence on the local environment, and the MEAM potential derived with the modified embedded atom method for GaN [32].

## 3. Results and discussion

### *3.1 Selection of the target effective charge and interatomic potential models*

To compare the accuracy of the effective target charge models, a comprehensive Monte Carlo simulation was made to analyze the relaxation of the electronic system. The simulation results for the 930 MeV Pb ion impacting GaN are illustrated in Fig. 3.

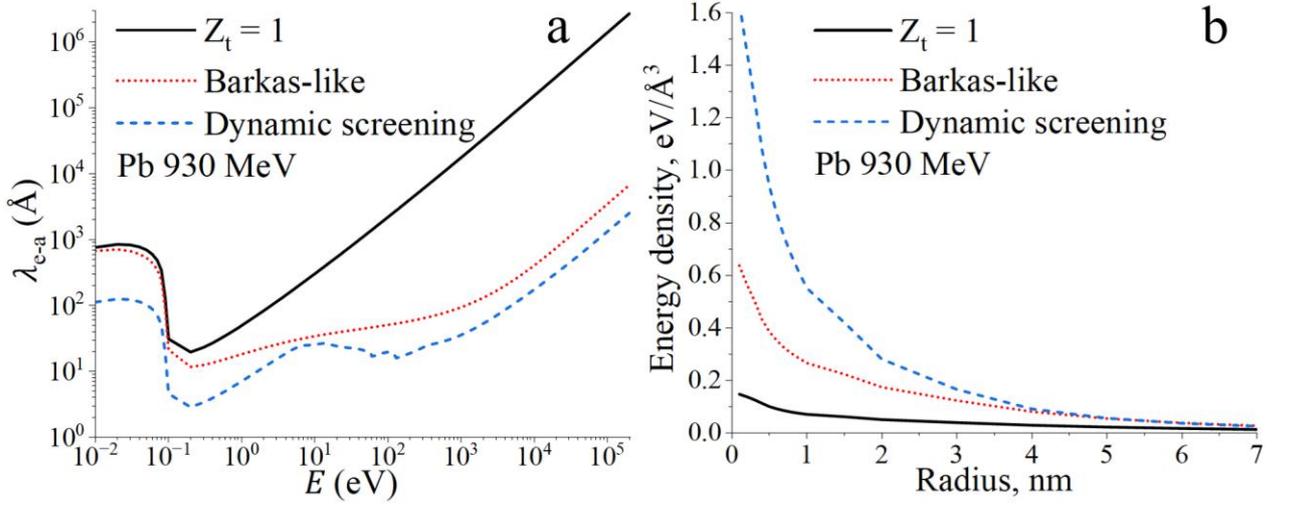

*Fig. 3 Comparison of (a) electrons mean free paths in atomic system of GaN and (b) transferred energy to the lattice for different $Z_t$.*

The electron elastic mean free path graph in the DS effective charge model shows a unique feature compared to other approximations. It exhibits a peak around 10 eV. This peak is attributed to a change in atomic screening behavior. As the incident electron energy surpasses the valence band energy of GaN (~20 eV), the screening effect by valence electrons diminishes. This leads to an increase in scattering cross-section, resulting in a decrease in the mean free path of electrons. A similar pattern is observed for SiC in Ref. [20].

Incorporating dynamical screening yields significantly higher values of energy transferred to the atomic lattice (an order of magnitude higher than for $Z_t = 1$). At the same time, the values of the elastic mean free path in this model for the low-energy region (<10 eV) are lower (the cross sections are larger) than the corresponding values calculated with the Barkas-like formula and $Z_t = 1$. In the high-energy region, the values obtained from the Barkas-like model significantly differ from the $Z_t = 1$ values and get closer to the DS model values.

The results of MD modeling of structural changes in tracks of 930 MeV Pb for various interatomic potentials (SW, TB, MEAM) and effective charge models ($Z_t = 1$, Barkas-like, DS) were compared with transmission electron microscopy data from Ref. [4].

The $Z_t = 1$ charge model results in minimal energy transfer into the lattice and does not cause any noticeable disruptions in the crystal structure when used in pair with any interatomic potential. Therefore, it was concluded that the $Z_t = 1$ charge model should be discarded.

Fig. 4 displays the results of MD simulations with cross-sections based on the dynamical screening target charge model corresponding to maximal excess lattice energy, and all three interatomic potentials. The SHI track radii values are also shown.

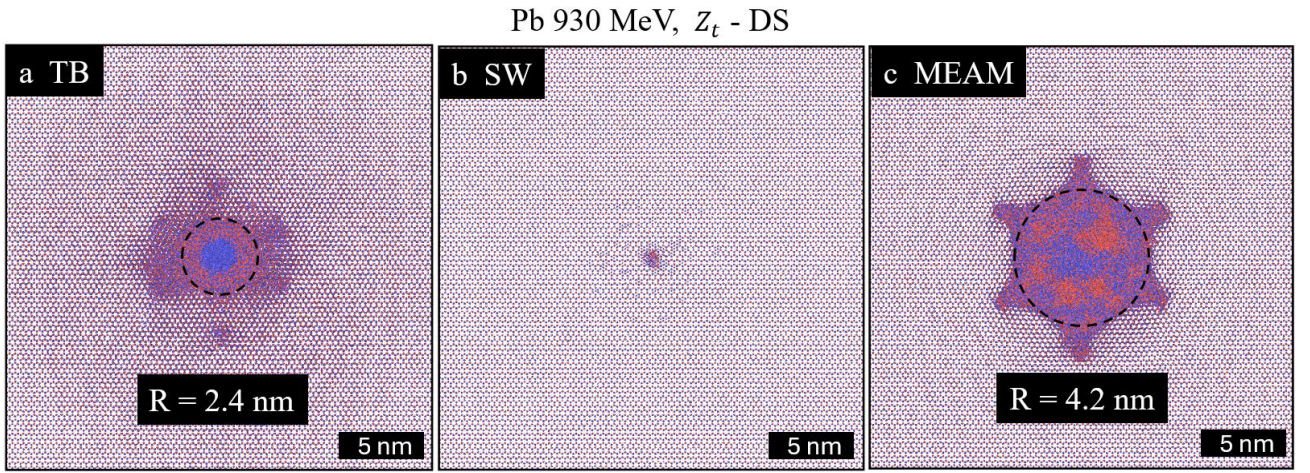

*Fig. 4 Track radii of 930 MeV Pb in GaN calculated within DS approximation using (a) Tersoff-Brenner, (b) Stillenger-Weber, (c) MEAM potentials.*

The Tersoff-Brenner potential predicts a track radius of 2.4 nm, matching experimental data (see Fig. 3 in Ref. [4]). In contrast, the application of the Stillinger-Weber potential results in complete recrystallization of the track with residual defects along the SHI trajectory. The other target atom charge models give lower deposited lattice energy values, providing no ion tracks in GaN when used in pair with SW potential. On the other hand, the MEAM potential overestimates the track radius by a factor of 2 compared to the experimental value. Thus, we decided to use the MEAM potential in pair with the Barkas-like charge model and compare it with the DS+TB approximation.

The Barkas-like formula with TB potential results in complete recrystallization of the track, while a track is formed when using the MEAM potential. Fig. 5 compares the results of applications of the most effective approaches: DS+TB and BL+MEAM.

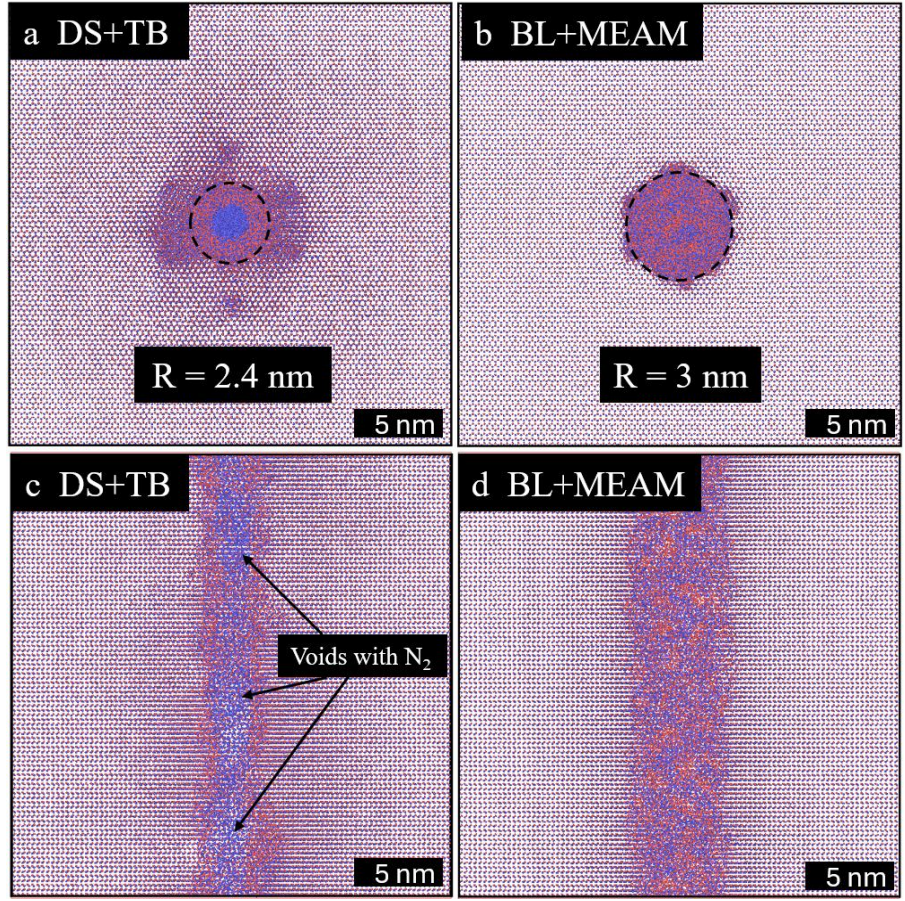

*Fig. 5 (a), (b) Track sizes and (c), (d) track structures with DS+TB and BL+MEAM approaches.*

Barkas+MEAM gives a track radius of 3 nm, which is close to the experimental value of 2.4 nm. However, the TEM images (see Figs. 4e,f in Ref. [6]) show that track cores contain voids. This result is only reproduced by the DS+TB approach.

Therefore, it seems that the DS effective charge model of target atoms coupled with the Tersoff-Brenner interatomic potential is the most appropriate approach for simulating of SHI track formation in GaN.

### *3.2 Track radii and structures*

Figs. 6 and 7 illustrate the results of our simulations of impacts of Au 185 MeV, Pb 132 MeV, Pb 930 MeV, and U 900 MeV ions in GaN. Appendix contains details on the electron kinetics and energy transferred to the atomic lattice in tracks of these ions. The forms and structures of tracks predicted by MD simulations were compared with those from the experiments [3-6] as well as the results of the modeling that couples thermal spike model with molecular dynamic simulations (TTM+MD) [5, 6], where the fitting procedure allowed the authors to replicate the experimental data.

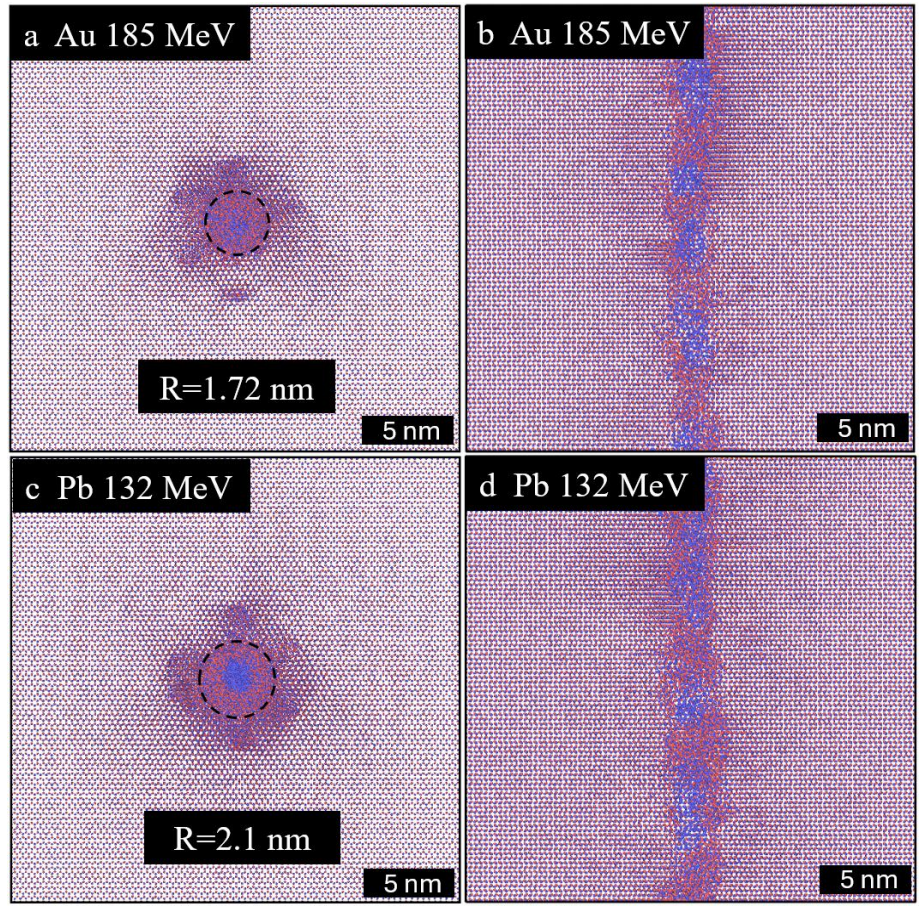

*Fig. 6 Radii and structures of (a) and (b) Au 185 MeV as well as (c) and (d) Pb 132 MeV tracks in GaN.*

Figs. 6a,b display the results of the simulation of an impact of 185 MeV Au ions into GaN. The ion tracks contain a cylindrical amorphous core surrounded by a damaged crystalline phase (track halo). The obtained radius of the amorphous core of 1.72 nm is consistent with both experimental and TTM+MD results (1.8 nm, see Figs. 1b, 2b in Ref. [5]). Notably, the calculations predict the presence of voids in track cores structure, which also were observed both in the TEM images and TTM+MD simulation (see Figs. 3a,c in Ref. [5]).

Figs. 6c,d show that the calculated in our simulations radius of the track core of 132 MeV Pb ion (2.1 nm) deviates by 0.6 nm from the experimental one (1.5 nm, see Fig. 1d in Ref. [3]) and by 0.8 nm from the radius predicted by TTM+MD model (1.3. nm, see Fig. 3b in Ref. [6]). This core contains nitrogen bubbles, similar to those forming in tracks of gold ions. Both our and the TTM+MD simulations demonstrated appearance of bubbles along the ion trajectory.

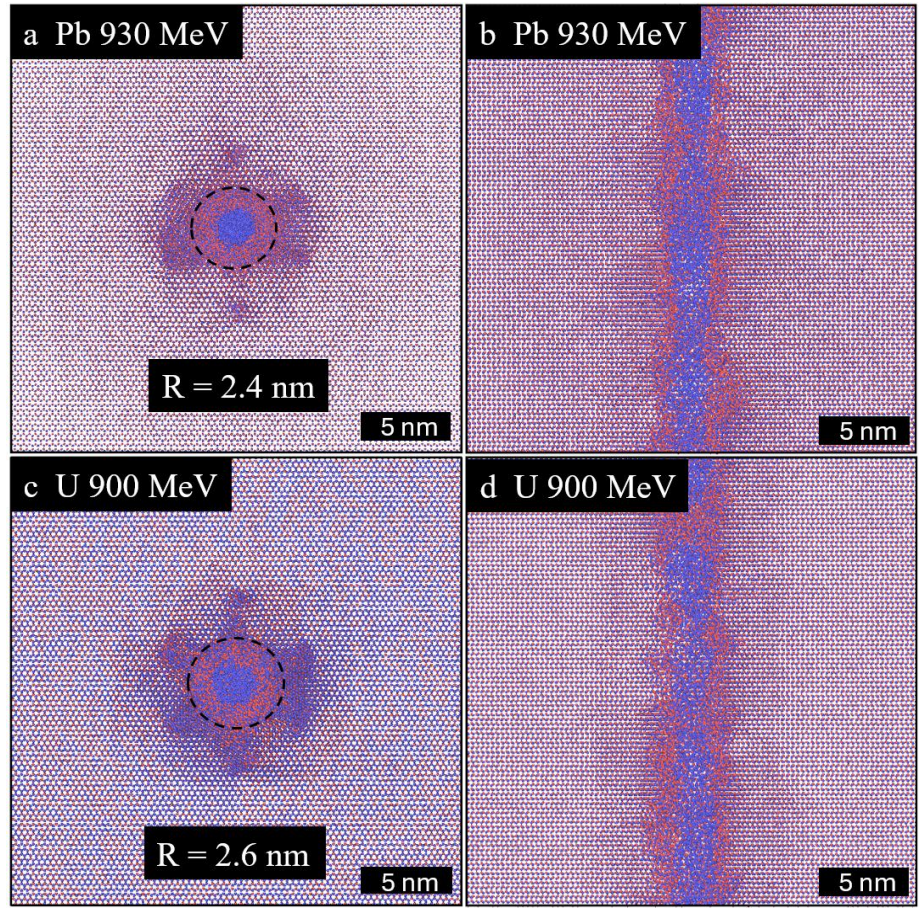

*Fig. 7 Radii and structure of (a) and (b) Pb 930 MeV as well as (c) and (d) U 900 MeV tracks in GaN.*

Figs. 7a,b present the results of the simulations of an impact of a 930 MeV Pb ion in GaN. The radius of the track obtained from our MD modeling matches with the experimental data (2.4 nm, see Fig. 3 in Ref. [4]), however it is larger than that obtained in TTM+MD simulations (2.2 nm, see Fig. 2b in Ref. [6]). Both the TREKIS+MD and TTM+MD modeling indicated the presence of voids in these tracks, consistent with the experimental results [6].

Figs. 7c,d illustrate the results of the simulations of an impact of a 900 MeV U ion in GaN. The TREKIS+MD model predicts the track core radius of 2.6 nm, deviating by 0.3 nm from the TTM+MD model and experimental core radii (2.9 nm, see Figs. 5b,d in Ref. [6]). The track core contains large nitrogen ($N_2$) filled voids similar to those obtained within TTM+MD simulations.

*3.3 Defects surrounding the track core*

The simulations demonstrate that at times less than 10 ps after the SHI impact, the damaged ion track initially appears as an amorphous structure. Subsequently, this highly damaged region shrinks due to recrystallization processes, and within ~100 ps, a heavily distorted crystalline structure forms around the core of the track.

To study this lattice damage, we applied dislocation extraction algorithm (DXA) [33] implemented in OVITO [27]. Fig. 8 illustrates evolution of the emerging defective structure during the relaxation of the lattice after the passage of a 930 MeV Pb ion in GaN. Atoms forming a wurtzite structure are marked in orange, while unidentified atoms, associated with the amorphous structure, are white. The reduction in the size of the amorphous phase occurs in the first 100 ps, and no visible changes observed further. The initial shape of this track is hexagonal, matching the crystal lattice symmetry.

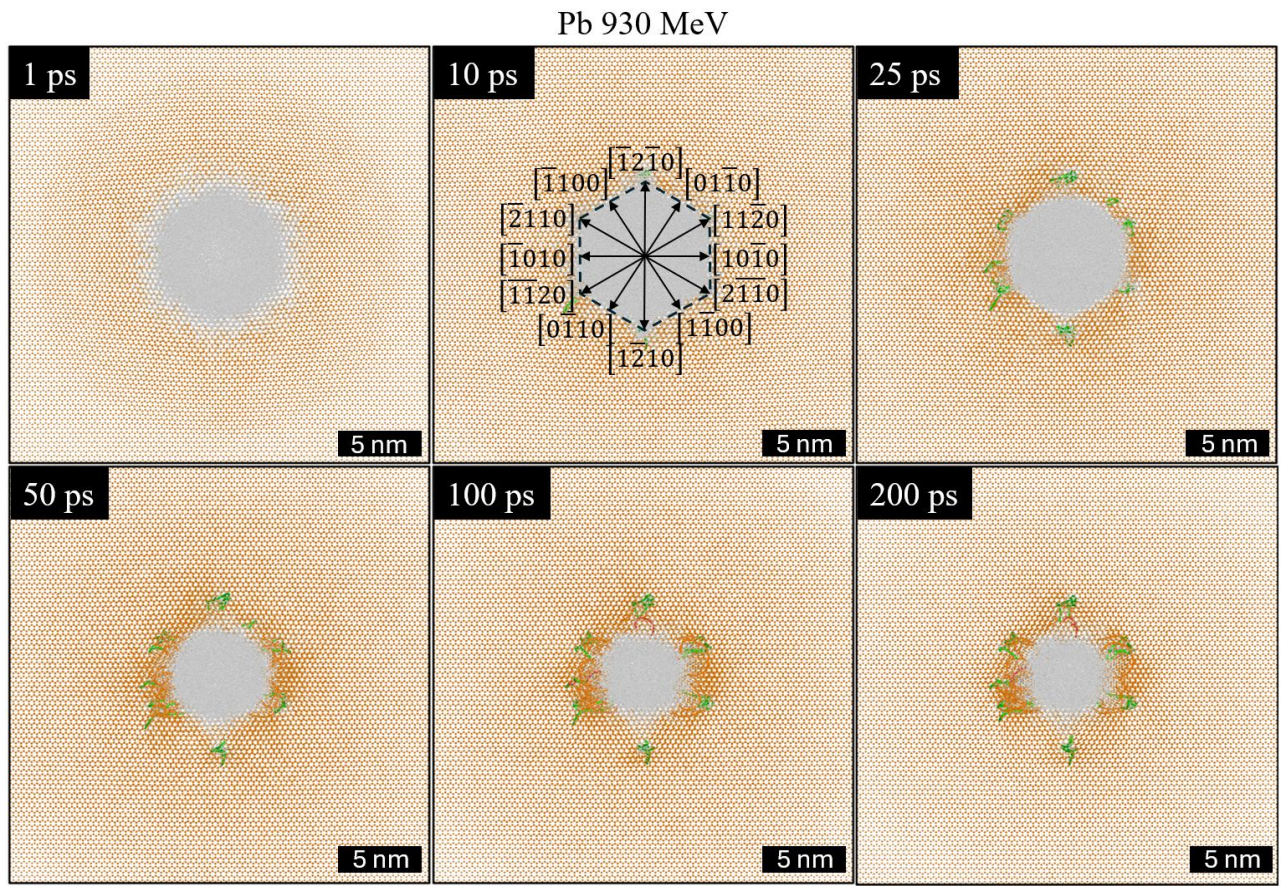

*Fig. 8 Evolution of dislocation structure in a 930 MeV Pb track in GaN; edge dislocations are shown as green lines, Shockley partial dislocations are shown as orange lines.*

Figs. 8 and 9 show the dislocations forming at the hexagon's vertices during the initial stages of recrystallization. Edge dislocations are shown as green lines, while Shockley partial dislocations, which were associated with stacking faults in Ref. [5], are shown as orange lines. The Burgers vectors of edge dislocations align with directions $[\bar{1}010]$, $[\bar{1}100]$, and $[01\bar{1}0]$. Similar defects were observed in tracks of all the SHI investigated in this study. The alignment of edge dislocations along the track may impede the recrystallization process. The TTM+MD model also predicts the formation of dislocations around the core of the 185 MeV Au track in GaN [5].

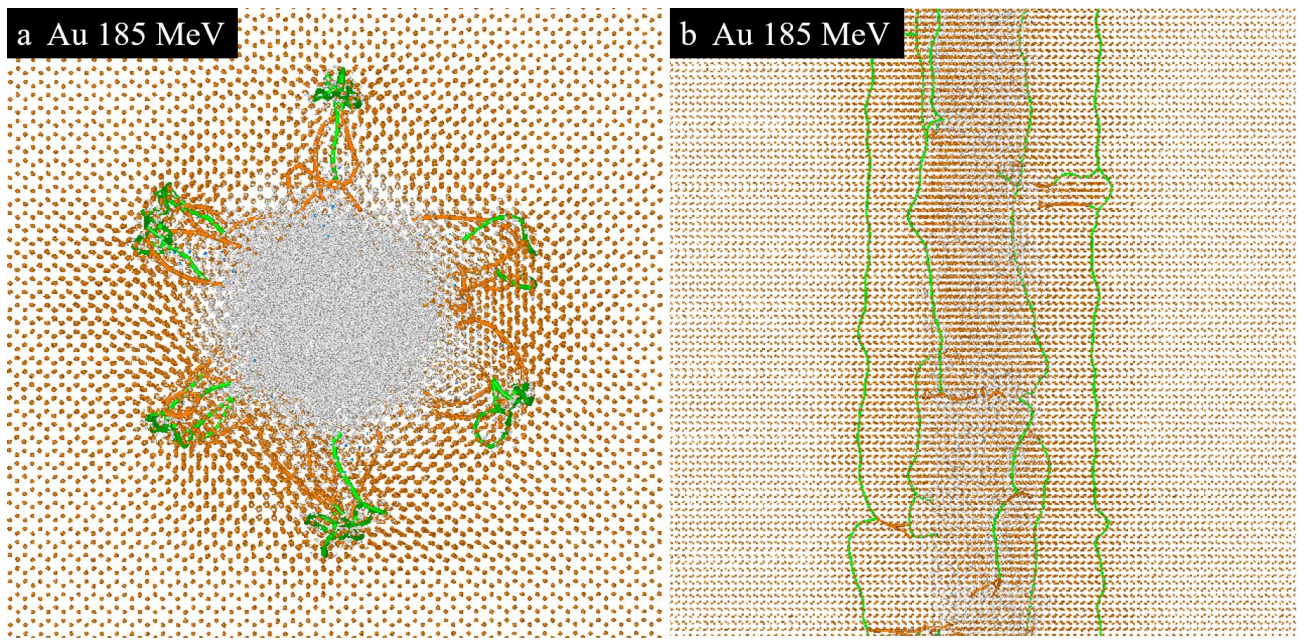

*Fig. 9 Lattice defects around the core of a 185 MeV Au track in GaN: (a) top view and (b) side view; edge dislocations are formed at the hexagon's vertices.*

**4. Conclusion**

The formation of damaged tracks in GaN was modeled with the multiscale TREKIS+MD model for different ions of different energies decelerating in the electron stopping regime: Au 185 MeV, Pb 132 MeV, Pb 930 MeV, and U 900 MeV.

The simulation results indicate the effect of the dynamical screening of target atoms during the scattering of generated fast electrons on structure transformations in nanometer proximity of SHI trajectories in this material.

The simulated tracks exhibited an amorphous phase in a nanoscale core surrounded by a heavily damaged crystalline region (halo). The core radii closely matched the experimental data. Nitrogen-filled voids (bubbles) were formed in the track cores, consistent with the TTM+MD modelling. The void formation was also observed in the experiments.

The defective crystalline lattice surrounding the track core contains edge dislocations at the corners of the damaged hexagon, which outlines the primary shape of the amorphous track before recrystallization. It is believed that these elongated edge dislocations are responsible for impeding the recrystallization of the material.

## Acknowledgements

This work has been carried out using computing resources of the federal collective usage center Complex for Simulation and Data Processing for Mega-science Facilities at NRC "Kurchatov Institute", http://ckp.nrcki.ru/.

This research was funded by the Science Committee of the Ministry of Science and Higher Education of the Republic of Kazakhstan (Grant No. AP23484126).

## Appendix

Fig. A1 illustrates the radial distributions of excited electrons in Pb 930 MeV track in GaN at various times after the ion passage. The graph depicts the propagation of two fronts. The first front represents the ballistic movement of fastest electrons through the undisturbed material. The second front occurs due to decay of plasmons appearing in the valence band of the material. At longer times diffusion plateau in the electron distribution appears resulting from deceleration of electrons in the medium.

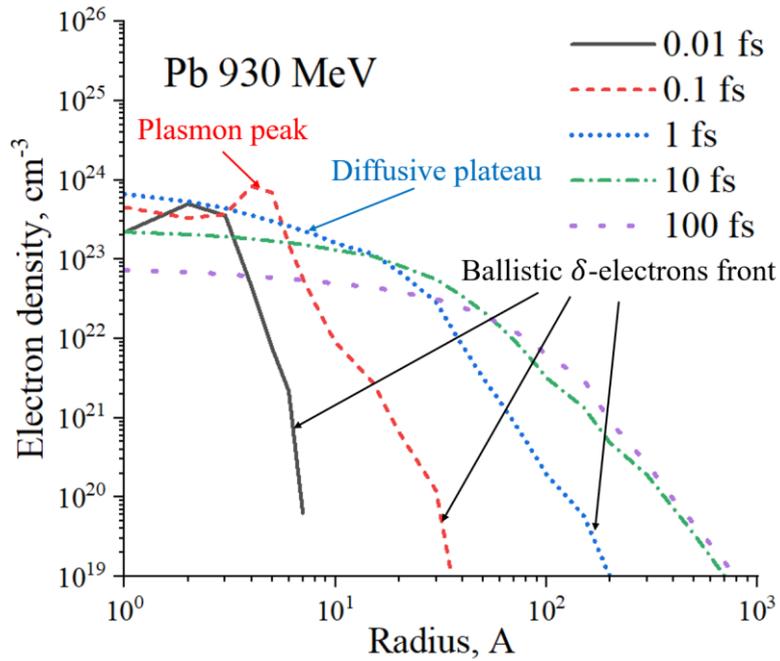

*Fig. A1 Radial distribution of excited electrons around the Pb 930 MeV trajectory at different times in GaN.*

The MC-simulations for 132 MeV Pb, 185 MeV Au, and 900 MeV U ions demonstrate similar results.

Fig. A2 presents the radial distributions of the energy transferred into GaN lattice at times 100 fs after impacts of these ions around their trajectories.

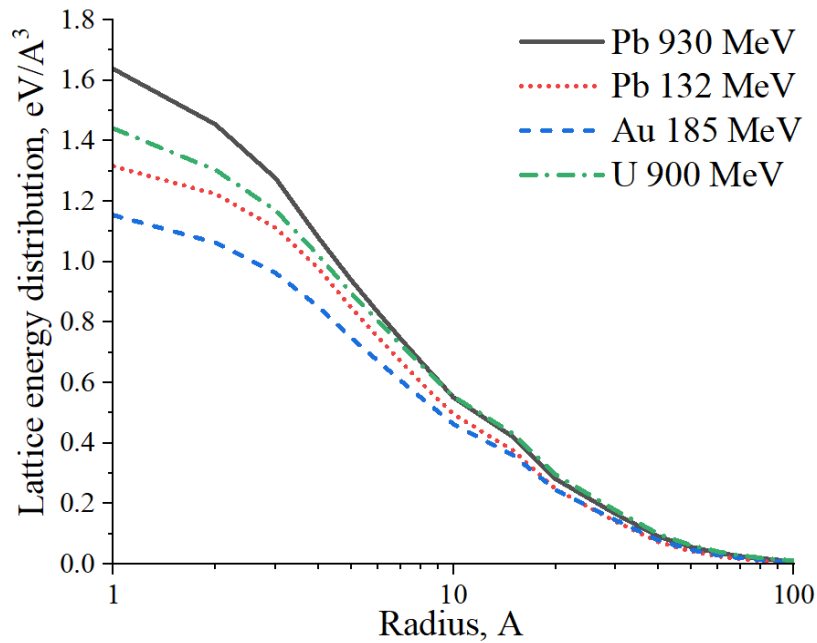

*Fig. A2 Radial distribution of the energy transferred to atoms of GaN at 100 fs after impacts of different ions round their trajectories.*